\documentclass[aps,prd,showpacs,twocolumn,altafillletter,superscriptaddress]{revtex4}

\usepackage{graphicx}
\usepackage{dcolumn}

\usepackage{hyperref} \usepackage{amssymb} \usepackage{amsmath}
\usepackage{color} \usepackage{epsfig} \usepackage{latexsym}
\usepackage{wasysym} \usepackage{comment} \usepackage{psfrag}
\usepackage[sort&compress]{natbib}

\newcommand{\gw}{gravitational wave } \newcommand{\gws}{gravitational waves }

\newcommand{\detodds}{{\mathcal O}_{(+,-)}}

\newcommand{\ee}[1]{\!\times\!10^{#1}} 
\newcommand{\prob}{{\rm P}}
\newcommand{\half}{\frac{1}{2}}

\newcommand{\dist}{{\mathcal D}}

\newcommand{\glitchepochGPS}{839457339}
\newcommand{\onsourcestart}{839457279}
\newcommand{\onsourceend}{839457399}

\newcommand{\preoffstart}{839447317}
\newcommand{\preoffend}{839457279}

\newcommand{\postoffstart}{839457399}
\newcommand{\postoffend}{839466903}

\newcommand{\preoffduration}{9962}
\newcommand{\postoffduration}{9504}

\newcommand{\onodds}{-5.03}
\newcommand{\FAP}{0.92}

\newcommand{\maxoffodds}{1.07}
\newcommand{\minoffodds}{-11.26}

\newcommand{\htwozero}{1.4 \ee{-20}}
\newcommand{\Etwozero}{5.0 \ee{44}}

\newcommand{\htwoone}{1.2 \ee{-20}}
\newcommand{\Etwoone}{1.3 \ee{45}}

\newcommand{\htwotwo}{6.3 \ee{-21}}
\newcommand{\Etwotwo}{6.3 \ee{44}}

\def\version$#1,v #2 #3${#2}

\makeatletter

\begin{document}


\title{A search for gravitational waves associated with the August 2006 timing
glitch of the Vela pulsar}

\input{authorlist}

\date{\today}



\begin{abstract}
The physical mechanisms responsible for pulsar timing glitches are thought to
excite quasi-normal mode oscillations in their parent neutron star that couple
to gravitational wave emission.  In August 2006, a timing glitch was observed
in the radio emission of PSR~B0833$-$45, the Vela pulsar.  At the time of the
glitch, the two co-located Hanford \gw detectors of the Laser Interferometer
Gravitational-wave observatory (LIGO) were operational and taking data as part of
the fifth LIGO science run (S5).  We present the first direct search for the
gravitational wave emission associated with oscillations of the fundamental
quadrupole mode excited by a pulsar timing glitch.  No \gw detection candidate
was found.  We place Bayesian $90\%$ confidence upper limits of $\htwotwo$ to
$\htwozero$ on the peak intrinsic strain amplitude of \gw ring-down
signals, depending on which spherical harmonic mode is excited. The corresponding
range of energy upper limits is $\Etwozero$ to $\Etwoone$\,erg.
\end{abstract}

\pacs{04.80.Nn, 07.05.Kf, 95.85.Sz, 97.60.Gb}

\maketitle

\section{Introduction}\label{sec:intro}
Neutron stars are often regarded as a prime source of various forms of
gravitational wave emission.  Recent searches for \gw emission from neutron
star systems include the search for the continuous, near-monochromatic
emission from rapidly rotating deformed neutron stars~\citep{S5knownPulsars}
and the characteristic chirp signal associated with the coalescence of a
binary neutron star or neutron star-black hole system~\citep{S5CBC,CBCGRB}.  An
additional mechanism for the radiation of gravitational waves from neutron
stars is the excitation of quasi-normal modes (QNMs) (see, for
example,~\citep{thorne1,thorne2,thorne3,thorne4,thorne5,thorne6,Kokkotas:LivingReview,SedrakianWhiteDwarfs}
and the references therein).  This excitation could occur as a consequence of
flaring activity in soft-gamma
repeaters~\citep{Pacheco:1998,SGRsearch,StackSearch}, the formation of a
hyper-massive neutron star following the coalescence of a binary neutron star
system~\citep{OechslinJanka}, or be associated with a pulsar timing glitch
caused by a star-quake or transfer of angular momentum from a superfluid core to a
solid crust~\citep{MiddleditchStarQuake,AndersonItoh}.

In this paper, we report the results of a search in data from the fifth science
run (S5) of the Laser Interferometer Gravitational-wave Observatory (LIGO) for
a gravitational wave signal produced by QNM excitation associated with a timing
glitch in the Vela pulsar in August 2006.  In Sec.~\ref{sec:VelaGlitch}, we briefly describe the
radio observations of the timing glitch that motivates this search and the
status of the LIGO gravitational wave detectors.  In
Sec.~\ref{sec:theory}, we describe the phenomenon of pulsar glitches and the
expected gravitational wave emission.  Section~\ref{sec:algorithm} describes
the details of the signal we search for and the Bayesian model selection
algorithm used for the analysis. Section~\ref{sec:results} reports the results
of the gravitational wave search.  Characterization of the sensitivity of the
search is described in Sec.~\ref{sec:validation}.  In
Sec.~\ref{sec:discussion}, we discuss these results and the prospects for
future searches.

\section{A glitch in PSR~B0833$-$45}\label{sec:VelaGlitch}

\subsection{Electromagnetic observations}\label{sec:Electromagnetic
Observations} PSR~B0833$-$45, known colloquially as the Vela pulsar, is
monitored almost daily by the Hartebeesthoek radio observatory (HartRAO) in
South Africa.  HartRAO performed three observations per day at 1668\,MHz and
2272\,MHz using a 26\,m telescope in a monitoring program that ran from 1985 to
2008~\citep{Buchner:2008}.  The radio pulse arrival times collected
by HartRAO indicate that a sudden increase in rotational frequency, a
phenomenon known as a pulsar glitch, occurred on August $12^{\rm th}$ 2006.

Following~\citep{ShemarLyne:1996}, observations of pulse arrival times from a
pulsar can be converted to rotational (angular) frequency residuals $\Delta \Omega$ relative to a
simple pre-glitch spin-down model of the form
\begin{equation}\label{eq:pre glitch phase}
\Omega(t) = \Omega_0 + \dot{\Omega}t,
\end{equation}
where $\Omega_0$ is the spin frequency at some reference time $t_0$ and
$\dot{\Omega}$ is its time derivative.  The post-glitch evolution of these
frequency residuals can be described as a permanent change in rotational
frequency $\Delta \Omega_p$ and its first and second derivatives $\Delta
\dot{\Omega}_p$ and $\Delta\ddot{\Omega}_p$, plus one or more transient
components which decay exponentially on a time-scale $\tau_i$ and have
amplitude $\Delta \Omega_i$. At time $t$, the residuals between the frequency
of pulses expected from the model in Eq.~(\ref{eq:pre glitch phase}) and those
which are observed following a glitch are then,
\begin{equation}\label{eq:post glitch phase}
\Delta \Omega (t) = \Delta \Omega_p + \Delta \dot{\Omega}_p t + \frac{1}{2} \Delta
\ddot{\Omega}_p t^2 + \sum_{i=1}^N \Delta \Omega_i e^{ -t/\tau_i},
\end{equation}
For this analysis we determined the  glitch epoch  by splitting the HartRAO
observations into pre- and post-glitch data sets.  Equation~(\ref{eq:pre glitch
phase}) was used to model 10 days of pre-glitch data.  Shorter lengths of
post-glitch data (2, 3 and 4 days) were then used to determine appropriate
post-glitch decay time-scales in Eq.~(\ref{eq:post glitch phase}) for this
event.  This yields a model for the post-glitch frequency residual evolution.
These pre- and post-glitch models were fitted to the HartRAO data using the
{\tt TEMPO2} phase-fitting software~\citep{TEMPO2}.  The intersection of these
models then determines the glitch epoch.  

We find that the glitch epoch is MJD $53959.9392 \pm 0.0002$
in terms of barycentric dynamical time at the solar system barycenter (UTC
2006--08--12 22:31:22 $\pm 17$, at the center of the Earth).  The analysis
presented in this work assumes the \gw emission is coincident in time with the
reported glitch epoch and uses 120\,seconds of data centered on the glitch
epoch corresponding to a timing uncertainty of greater than $3\mbox{-}\sigma$.

The magnitude of the glitch, relative to the pre-glitch rotational frequency of
$\Omega_0 \approx 2\pi \times 11$\, rads$^{-1}$, was $\Delta \Omega/\Omega_0 =
2.620 \times 10^{-6}$~\citep{Flanagan:2006}.   For comparison, the largest
glitch observed to date in the Vela pulsar had magnitude $\Delta \Omega /
\Omega_0 = 3.1\times 10^{-6}$~\cite{BigVelaGlitch}.

As well as the radio observations of the glitch in PSR~B0833$-$45, our \gw
search makes use of Chandra X-ray telescope observations which determine the
spin inclination $\iota$ and position angle $\psi_G$.  The inclination is the
angle between the pulsar's rotation axis and the line of sight to the Earth.
The position angle is the angle between Celestial North and the spin axis,
counter-clockwise in the plane of the sky~\citep{Ng:2008}.  Finally, Hubble
Space Telescope observations of parallax indicate that Vela is a particularly
nearby radio pulsar at a distance of just
$287^{+19}_{-17}$\,pc~\citep{Dodson:2003}.  Table~\ref{table:vela parameters}
gives a summary of parameters specific to the Vela pulsar and the August 2006
glitch.  Further details and measurements can be found in the ATNF pulsar
catalogue~\citep{ATNF:website, ATNF:paper}.

\begin{table}
\begin{center}
\begin{tabular}{l c c}
\hline PSR B0833$-$45 & \\
\hline \hline
Right ascension\footnotemark[1] &   $\alpha$  & $08^{\rm h} 35^{\rm m} 20.61149''$ \\ 
Declination\footnotemark[1]             &   $\delta$  & $-45^{\circ} 10'34.8751''$  \\
Spin inclination\footnotemark[2]     &   $\iota$   & $63.60^{+0.07}_{-0.05}\pm1.3^{\circ}$ \\
Polarization Angle         &   $\psi_G$ & $130.63^{+0.05}_{-0.07}$ \\
Glitch epoch     &   $T_{\rm glitch}$   & MJD $53959.9392 \pm 0.0002$ \\
 			     &   			  	    & GPS $\glitchepochGPS \pm 17$ \\
				 & 						& UTC 2006--08--12 22:31:22 $\pm 17$\\ 
Spin frequency          &   $\Omega_0/2\pi$ & $11.191455227602 \pm 1.8\times10^{-11}$\,Hz        \\
Frequency epoch &  & MJD 53945 \\
Fractional glitch size\footnotemark[3]  &   $\Delta \Omega / \Omega_0$ & $2.620\times10^{-6}$\\
Distance\footnotemark[4] &   $\dist$     &       $287^{+19}_{-17}$\,pc \\ 
\hline \end{tabular}
\footnotetext[1]{Taken from~\citep{ATNF:website,ATNF:paper}}
\footnotetext[2]{Taken from~\citep{Ng:2008} }
\footnotetext[3]{Taken from~\citep{Flanagan:2006} }
\footnotetext[4]{Taken from~\citep{Dodson:2003} }
\caption[PSR B0833$-$45 parameters]{\label{table:vela parameters}
Parameters of the Vela pulsar.  The statistical
and systematic errors in $\iota$ are listed as the first and second terms,
respectively.  The spin frequency and the glitch epoch were determined from the
analysis described in Sec.~\ref{sec:Electromagnetic Observations}.  The error
in the glitch epoch is an estimate of the $1\mbox{-}\sigma$ uncertainty.  The
glitch epoch quoted as MJD is defined in terms of barycentric dynamical time at
the solar system barycenter.  GPS and UTC times are terrestrial. The frequency
epoch is the epoch at which the pre-glitch spin frequency was estimated.}
\end{center}
\end{table}

\subsection{LIGO data}\label{sec:ligo data}
At the time of the Vela glitch, LIGO was
operating three laser interferometric detectors at two observatories
in the United States. Two detectors were operating at the Hanford site, one
with 4\,km arms and another with 2\,km arms.  These are
referred to as H1 and H2, respectively.  A third detector, with 4\,km
arms, was operating at the Livingston site, referred to as L1.  A full
description of the configuration and status of the LIGO detectors during S5
can
be found in~\citep{LIGO}.  There are no data from either the GE0~600 or Virgo
\gw detectors which cover the glitch epoch.

The data from the two Hanford detectors around the time of the pulsar glitch are
of very high quality and completely contiguous for a time window centered on
the glitch epoch lasting nearly five and a half hours.  The Livingston detector
was operating at the time of the glitch, but began to suffer from a degradation
in data quality due to elevated seismic noise approximately thirty
seconds later, and lost lock (the resonance condition of the Fabry-Perot arm
cavities) less than three minutes after that.  We have therefore chosen not to
include L1 data in this analysis due to the instability of the detector during
this period and the reduction in the amount of off-source data available (see
Sec.~\ref{sec:algorithm}).  In GPS time, the glitch epoch is
\glitchepochGPS $\pm$17.  There are 19586 seconds of data available from H1 and H2 in
the period $[839447317, 839466903)$ before H1 and H2 also begin to suffer from
degradations in data quality.  This entire contiguous segment is used in the
analysis.

\section{Pulsar glitches \& gravitational radiation}\label{sec:theory}
The physical mechanism behind pulsar glitches is not known.
It is not even known if all glitches are caused by the same mechanism.
Currently most theories fall into two classes: crust fracture (``star-quakes'')
and superfluid-crust interactions.  These produce different estimates of the
maximum energy and gravitational-wave strain to be expected.

The magnitudes of glitches in the Vela pulsar and the
frequency with which they occur are indicative of being driven by the
interaction of an internal superfluid with the solid crust of the neutron
star~\citep{AnderssonComerPrix}. For these superfluid-driven glitches, there
may be a series of incoherent, band-limited bursts of \gws due to an avalanche
of vortex rearrangements~\citep{MelatosAvalancheDynamics}.  This signal is
predicted to
occur during the rise-time of the glitch ($\leq 40$\,seconds before the observed jump
in frequency).  A possible consequence of this vortex avalanche is the
excitation of one or more of the families of global oscillations in the neutron
star.  These families are divided according to their respective restoring
forces (e.g., the fundamental (\emph{f}) modes, pressure (\emph{p}) modes,
buoyancy (\emph{g}) modes and space-time (\emph{w}) modes)~\citep{Sidery}.
These oscillations will be at least partially damped by gravitational wave
emission on timescales of milliseconds to seconds, leading to a characteristic
gravitational wave signal in the form of a decaying sinusoid.  There may also
be a continuous periodic signal near the spin frequency of the star due to
non-axisymmetric Ekman flow~\citep{MelatosGravRad}. This emission dies away on
the same time-scale as the post-glitch recovery of the pulsar spin
frequency ($\sim 14$ days).

Alternatively, the glitch may have been caused by a star-quake due to a
spin-down induced relaxation of ellipticity~\citep{Ruderman:1969}, although
the size and rate of the glitches mean that this cannot explain all of
them~\cite{BaymPines}.
In this
case, it seems likely that oscillation modes will also be excited. The amount
of excitation of the various mode families is not clear and will depend
on the internal dynamics of the star during the quake.

Due to the gravitational-wave damping rates of the various mode families, it is reasonable to assume that the bulk of \gw emission associated with
oscillatory motion is generated by mass quadrupole (i.e. spherical harmonic
index $l=2$) $f$-mode oscillations.  Furthermore, we make the simplifying
assumption that a single harmonic dominates, so that the \gw emission from the
$f$-mode oscillations can be characterized entirely by the harmonic indices
$l=2$ and one of the $2l+1$ values of $m$.  This assumption and its
astrophysical interpretation are discussed further in Sec.~\ref{sec:discussion}.
The plus ($+$) and cross ($\times$) polarizations for each spherical
harmonic mode in this model are:
\begin{widetext}
\begin{subequations}
\begin{eqnarray}\label{eq:hplus}
h^{2m}_+(t)  =  
\begin{cases}
h_{2m}{\cal A}_{+}^{2m} \sin[2\pi\nu_0 (t-t_0) + \delta_0]
e^{-(t-t_0)/\tau_0}~\text{for $t\geq t_0$}, \\
0~\text{otherwise.}
 \end{cases}
\end{eqnarray}

\begin{eqnarray}\label{eq:hcross}
h^{2m}_{\times}(t)  =  
\begin{cases}
h_{2m}{\cal A}_{\times}^{2m} \cos[2\pi\nu_0 (t-t_0) + \delta_0] e^{-(t-t_0)/\tau_0}~\text{for $t\geq t_0$}, \\
0~\text{otherwise.}
 \end{cases}
\end{eqnarray}
\end{subequations}
\end{widetext}
We refer to this decaying sinusoidal signal as a \emph{ring-down} with
frequency $\nu_0$, damping time $\tau_0$ and phase
$\delta_0$. The amplitude $h_{2m}$ is the \emph{peak intrinsic} \gw strain
emitted by any one of the various $l=2, m=-2,\dots,2$ modes.  The amplitude
terms ${\cal A}_{+,\times}^{2m}$ encode the angular dependence of the \gw
emission around the star for the $m^{\rm th}$ harmonic and depend on the
line-of-sight inclination angle $\iota$.  Their explicit dependencies can be
found in table~\ref{table:inclinations} and are calculated from tensor spherical
harmonics as in~\citep{Thorne:1980}.

\begin{table}[h!]
\begin{center}
\begin{tabular}{l c c } 
\hline 
Spherical Harmonic Indices & ${\mathcal A}^{2m}_+$  & ${\mathcal A}^{2m}_{\times}$
\\[0.15cm]\hline\hline
$l=2,~m=0$   & $\sin^2\iota$ & 0 \\ 
$l=2,~m=\pm1$ & $\sin 2\iota$ & $2\sin \iota$ \\ 
$l=2,~m=\pm2$ & $1+\cos^2 \iota$ & $2\cos \iota$
\end{tabular}
\caption[]{The line-of-sight inclination angle $\iota$ dependencies of the
expected polarizations in equations~\ref{eq:hplus} and~\ref{eq:hcross} for each
set of spherical harmonic indices ($l,~m$).\label{table:inclinations}}
\end{center}
\end{table}

The $f$-mode frequency and damping time are sensitive to the equation of
state of the neutron star, which is not known.
Calculations of the frequency and damping time of the fundamental
quadrupole mode for various models of the equation of state, such as those
in~\citep{Andersson:1998,Benhar:2005}, indicate that the frequency lies in the
range $1 \lesssim \nu_0 \lesssim 3$\,kHz and the damping time lies in the
range $0.05\lesssim\tau_0\lesssim 0.5$\,seconds.

If we assume that a change in rotational angular frequency of size $\Delta \Omega$ is
caused by a change in the moment of inertia, corresponding to a star-quake,
it can be shown that the resulting change in rotational energy is given by
$\Delta E =  \frac{1}{2}I_* \Omega \Delta \Omega$, where $I_*$ is the stellar
moment of inertia and we assume conservation of angular momentum.
Inserting fiducial values for the moment of inertia~\citep{LattimerPrakash:2000},
rotational velocity and pulsar glitch magnitude we see that the characteristic
energy associated with pulsar glitches driven by seismic activity is
\begin{eqnarray}\label{eq:quake energy}
\Delta E_{\rm quake} & \approx & 10^{42}\,{\rm erg}~
\left(\frac{I_*}{10^{38}\,\text{kg m}^2}\right)\nonumber\\ & \times &
\left(\frac{\Omega}{20\pi\,\text{rad s}^{-1}}\right)^2 \left(\frac{\Delta
\Omega / \Omega}{10^{-6}}\right), \end{eqnarray}
where we have used the spin-frequency of Vela and a glitch magnitude of
$\Delta \Omega/\Omega = 10^{-6}$, typical of those in Vela.
This is then the maximum energy that could be radiated in gravitational waves.

For superfluid-driven glitches, an alternative approach to computing the
characteristic energy is to directly compute the change in gravitational potential
energy resulting from the net loss of rotational kinetic energy in the context
of a two-stream instability model~\citep{AnderssonComerPrix}. In this picture,
there exists a critical difference in the rotational angular frequency between
a differentially rotating crust and superfluid interior.  Beyond this critical
\emph{lag} frequency $\Omega_{\rm lag}$, the superfluid interior suddenly and
dramatically couples to the solid crust.  During the glitch, a fraction of the
excess angular momentum in the superfluid is imparted to the crust so that the
superfluid spins down while the crust spins up. It can then be shown that
the change in the rotational energy is, to leading order,
$\Delta E \approx  -I_c \Omega^2 \left(\Delta \Omega / \Omega\right)
\left(\Omega_{\rm lag} / \Omega\right)$, where $I_c$ is the moment of inertia
of the solid crust only.
Inserting fiducial values, we find:
\begin{eqnarray}\label{eq:vortex energy}
\Delta E_{\rm vortex} & \approx & 10^{38}\,{\rm erg}~ \left(\frac{I_c}{10^{37}
\text{ kg m}^2}\right) \left(\frac{\Omega}{20\pi\,\text{rad s}^{-1}}\right)^2 \nonumber \\
& \times & 
\left(\frac{\Delta \Omega/\Omega}{10^{-6}}\right) \left(\frac{\Omega_{\rm
lag}/\Omega}{5\ee{-4}}\right),
\end{eqnarray}
where we have assumed $\Omega_{\rm lag} \sim 5\times 10^{-4} \Omega$~\citep{LyneShemarSmith} and we have
assumed that the moment of inertia of the crust is about $10\%$ of the total
stellar moment of inertia.

An estimate of the intrinsic peak amplitude of \gws emitted in the form of
ring-downs as described by Eq.~(\ref{eq:hplus}) and Eq.~(\ref{eq:hcross}) can be
found by integrating the luminosity of that signal over time and solid angle.
Assuming that all of the rotational energy released by the glitch goes into
exciting a single spherical harmonic and that the oscillations are completely damped by
\gw emission, we find that the expected peak amplitude of a ring-down signal
is

\begin{eqnarray}\label{eq:fiducial amplitude}
h_{2m} \approx & 10^{-23} &
\left(\frac{E_{2m}}{10^{42}\,{\rm erg}}\right)^{\half}
\left(\frac{2\,\text{kHz}}{\nu_0}\right) \nonumber \\
& \times & \left(\frac{200\,\text{ms}}{\tau_0}\right)^{\half}
\left(\frac{1\,\text{kpc}}{\dist}\right). 
\end{eqnarray}
%

\section{Bayesian model selection algorithm}\label{sec:algorithm} 
This search updates and deploys the model selection algorithm previously
described in~\citep{Clark:2007}.  Bayesian model selection is performed by
evaluating the ratio of the posterior probabilities between two competing
models describing the data.  Following the work
in~\citep{Clark:2007,Veitch:2008} and~\citep{Searle:2008}, let us suppose our
models represent some data $D$ which contains a \gw signal, called the
\emph{detection model}, denoted $M_+$, and data which does not contain a \gw
signal, called the \emph{null-detection model}, $M_-$. Writing out the ratio of
the posterior probabilities of each model, we see that
%
%
\begin{subequations}
\begin{eqnarray}\label{eq:posterior odds}
\detodds & = & \frac{\prob(M_{\rm +}|D)}{\prob(M_{-}|D)} \\
& = & \frac{\prob(M_{\rm +})}{\prob(M_{\rm -})}\frac{\prob(D|M_{\rm
+})}{\prob(D|M_{\rm -})},
\end{eqnarray} \end{subequations}
The first term is commonly referred to as the \emph{prior odds} and indicates
the ratio of belief one has in the competing models prior to performing the experiment.
Since it can be difficult to estimate, it is common to set this equal to unity.  The
second term, the \emph{Bayes factor}, is the ratio of the marginal likelihoods
or \emph{evidences} for the data, given each model.  For a model $M_i$
described by a set of parameters $\vec{\mu}$, the evidence is computed from
\begin{equation}\label{eq:evidence integral} 
\prob(D|M_i) = \int_{\mu} p(\vec{\mu}|M_i) p(D|\vec{\mu},M_i)~{\rm
d}\vec{\mu},
\end{equation}
where $p(\vec{\mu}|M_i)$ is the prior probability density distribution on the
parameters $\vec{\mu}$ and $p(D|\vec{\mu},M_i)$ is the likelihood
of obtaining the data $D$, given parameter values $\vec{\mu}$.  The details
of the models $M_+$ and $M_-$ as used in this analysis are given in Sec.~\ref{sec:signal model}.

\begin{figure}\centering \scalebox{0.6}{\includegraphics{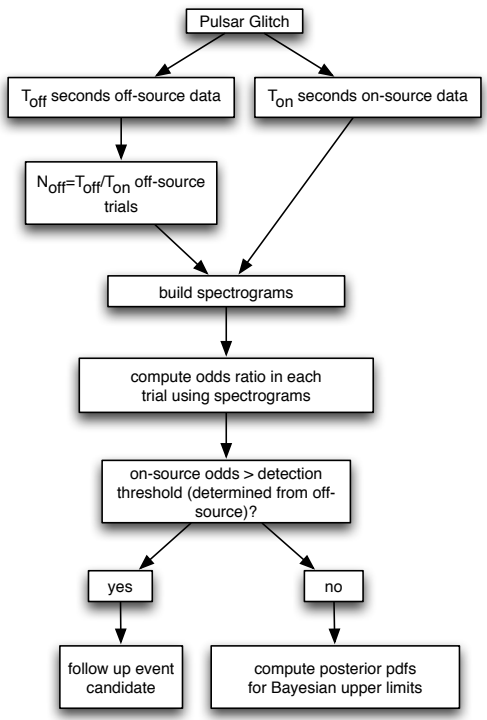}}
\caption[]{A schematic view of the analysis pipeline.  The odds ratio
$\detodds$ is evaluated using on and off-source data near the pulsar timing
glitch.  If the odds ratio in the on-source data is greater than that expected from
the distribution of odds ratios in in the off-source data, we have a
candidate event for follow-up investigations.  If there is no significant excess in $\detodds$ in the on-source
data, we obtain upper limits on the gravitational wave amplitude and
energy.\label{fig:pipeline}}
\end{figure}

The data analysis procedure is shown schematically in Fig.~\ref{fig:pipeline}.
Gravitational wave detector time series data centered on the pulsar glitch
epoch and spanning the uncertainty in the epoch is obtained.  This
constitutes the \emph{on-source} data and has duration $T_{\rm on}$ seconds.
We also obtain a longer segment of time series data from before and after the
on-source period.  This is termed \emph{off-source} data and is used to
estimate the distribution and behaviour of the detection statistic (in
our case, the odds ratio $\detodds$).  The off-source data has total duration
$T_{\rm off}$ seconds.   This off-source data is then further divided into
$N_{\rm off}=T_{\rm off}$/$T_{\rm on}$ trials, each of which will be used to
compute one value of the odds ratio.

The data from each detector in the one on-source trial and each of the
$N_{\rm off}$ off-source trials are then divided into short,
overlapping time segments and a high-pass $12^{\rm th}$ order Butterworth filter is
applied with a knee frequency of 800\,Hz.  The power spectral density
in that segment is then computed and we form a time-frequency map
of power, or \emph{spectrogram}, for each detector.  The parameters used to
construct the spectrograms are given in Table~\ref{table:search parameters}.
These spectrograms are then used as the data $D_1$ and $D_2$ from which we
compute the odds ratio $\detodds$.  Values of $\detodds \gg 1$ indicate a
significant preference for the detection model.

\begin{center}
\begin{table}[h!]
\begin{tabular}{l c c}
\hline
Parameter Space \\
\hline \hline 
On-source data (GPS) & $[\onsourcestart,~\onsourceend)$ \\
Off-source data (GPS) & $[\preoffstart,~\preoffend)$ \\
                & $[\postoffstart,~\postoffend)$ \\
LIGO antenna factors & $F_+=-0.69$, $F_{\times}=-0.15$ \\
Signal frequency ($\nu_0$) range & $[1,3]$\,kHz \\
Decay time ($\tau_0$) range & $[50,500]$\,ms \\ 
Amplitude (${\mathcal A}_{\rm eff}$) range  & $[10^{-22},10^{-19}]$ \\[0.1cm]
\hline
Spectrogram configuration & \\
\hline \hline
Fourier segment length & $2$\,seconds \\ 
Overlap  & $1.5$\,seconds \\ 
Frequency resolution & $0.5$\,Hz \\
Data sampling frequency & 16384\,Hz \\
\hline \end{tabular}
\caption[Search Parameters]{Parameters used in the gravitational wave data
analysis.  The antenna factors have been computed for the LIGO Hanford Observatory, the sky
location and polarization angle for Vela, and the time of the glitch
\label{table:search parameters}.} \end{table}
\end{center}

The LIGO detector noise is, in general, non-stationary and can be found to contain
instrumental or environmental transient signals which tend to mimic the \gw
signal we are looking for.  To mitigate the risk of falsely claiming a \gw
detection, the off-source data is used to empirically determine the
distribution of $\detodds$ when we do not expect a \gw signal to be present.
This allows us to estimate the statistical significance of any given value of
$\detodds$.  We then compare the value of the odds ratio computed from the
on-source data with this empirical distribution.  If the significance of the
on-source value of $\detodds$ is greater than the most significant
off-source value then we have an interesting event candidate which merits
further investigations such as a more robust estimate of its significance above the
background level and verification with other data analysis pipelines.  
In this sense then, although the detection statistic
itself, the odds ratio $\detodds$, is formed from Bayesian arguments, we choose
a frequentist interpretation of its significance due to our inability to
accurately model spurious instrumental noise features in the detector data.  If
no detection candidate is found, $90\%$ confidence upper limits on the
intrinsic gravitational wave strain amplitude $h_{2m}$ and energy $E_{2m}$ are
found from their respective posterior probability density functions.

\subsection{Signal model and computing the evidence for gravitational wave
detection}\label{sec:signal model} Recall that we consider the detection
and upper limits of each spherical harmonic mode (indexed by $l=2,~m$)
separately.  The response of an interferometric \gw detector to an impinging
gravitational wave is such that the time-domain signal in the detector output
can be written
\begin{equation}\label{eq:detector response}
s^{2m}(t) = F_+(\mathbf{\Theta},\psi_G)h^{2m}_+(t) +
F_{\times}(\mathbf{\Theta},\psi_G)h^{2m}_{\times}(t),
\end{equation}
where $h^{2m}_{+,\times}$ are given by Eqs.~(\ref{eq:hplus})
and~(\ref{eq:hcross}).  The terms $F_{+,\times}(\mathbf{\Theta},\psi)$ are the
detector response functions to the plus and cross polarizations of the
gravitational waves, defined in~\citep{JKS}.  These are functions of the sky
location of the source $\mathbf{\Theta}=\{\alpha,\delta\}$, and the \gw
polarization angle $\psi_G$.  We take the polarization angle to be equal to the
position angle defined in~\citep{Ng:2004}.
For a single detector location and short-duration signal, where the antenna
factors $F_{+,\times}$ are fixed, we are free to adopt a simplified signal
model and absorb all of the orientation factors ($F_{+,\times}$ and
${\mathcal A}_{+,\times}^{2m}$) into a single effective amplitude term ${\mathcal
A}_{\rm eff}$.  Our time-domain signal model is finally
\begin{equation}\label{eq:signal model}
s(t)  =  
\begin{cases}
{\mathcal A}_{\rm eff} \sin\left[2\pi \nu_0 (t-t_0) +
\delta_0^{'}\right] e^{-(t-t_0)/\tau_0}~\text{for $t_0\ge0$}, \\
0~\text{otherwise,}
 \end{cases}
\end{equation}
where the phase term $\delta_0^{'}$ is now primed since it has been affected
by the combination of the two signal polarizations into a single sinusoidal
component.  Note, however, that this analysis uses the power spectral density
of the data and is insensitive to the signal phase.  

We can then use the effective amplitude, the known inclination dependence
encoded in the ${\mathcal A}_+$ and ${\mathcal A}_{\times}$ terms for the
individual spherical harmonics, and the detector antenna factors $F_+$ and
$F_{\times}$ to convert the effective amplitude ${\mathcal A}_{\rm eff}$ to the
intrinsic \gw strain amplitude of the $m^{\rm th}$ mode, $h_{2m}$:
\begin{equation}\label{eq:effective amplitude}
h_{2m} = \frac{{\mathcal A}_{\rm eff}}{
\left[\left(F_+ {\mathcal A}_{+}^{2m}\right)^2 +\left(F_{\times}
{\mathcal A}_{\times}^{2m}\right)^2\right]^{\half}},
\end{equation}
which we note is insensitive to the sign of $m$.  Upper limits on \gw
amplitude and energy are later presented for each value of $|m|$.

The likelihood function, which describes the probability
of observing the power $\tilde{d}_{ij}$ in the ($i^{\rm th}$, $j^{\rm th}$)
spectrogram pixel (time, frequency) given an expected signal power $\tilde{s}_{ij}(\vec{\mu})$,
is a non-central $\chi^2$ distribution with two degrees of freedom and a
non-centrality parameter given by the expected contribution to the power from
the model whose likelihood we are evaluating.  For the case where a gravitational
wave signal parameterized by $\vec{\mu}$ contributes power
$\tilde{s}_{ij}(\vec{\mu})$, the joint likelihood for the entire spectrogram
is
\begin{eqnarray}\label{eq:signal likelihood} p(D|\vec{\mu},M_{\rm +}) & = &
\prod_{i=1}^{N_{\rm T}}\prod_{j=1}^{N_{\rm F}} \left\{
\frac{1}{2\sigma_j^2} \exp\left[-\frac{\tilde{d}_{ij} +
\tilde{s}_{ij}(\vec{\mu})}{2\sigma_j^2}\right] \right. \nonumber \\ & \times &
\left. I_0\left[\frac{1}{\sigma_j^2}\sqrt{\tilde{d}_{ij} \tilde{s}_{ij}(\vec{\mu})}\right]\right\}, \end{eqnarray}
where there are $N_T$ total time bins in the spectrogram, $N_F$ frequency bins
and $\sigma_j^2$ is the variance of the noise power in the $j^{\rm th}$
frequency bin.  $I_0$ is the zeroth order modified Bessel function.  The
noise power variance $\sigma_j^2$ is estimated from the median noise power
across time bins at that frequency using the data segment which is being
analyzed.  This method of estimating the noise is robust against bursts of
power shorter than the length of the on-source data and avoids
the potential contamination of the estimate of $\sigma_j^2$ from both instrumental noise
artifacts and gravitational wave signals.

The prior probability distributions on the ring-down frequency $\nu_0$ and
damping time $\tau_0$ are guided by the eigenmode calculations
in~\citep{Andersson:1998} and~\citep{Benhar:2005}.  The frequency prior is
taken to be uniform between 1 and 3\,kHz and the damping time prior uniform
between 50 and 500\,ms.  The glitch epoch for the search described here is
found to have a $1\mbox{-}\sigma$ uncertainty of 17\,seconds.  We adopt a
conservative flat prior range on the start time of the signal $t_0$ with a
total width of 120\,seconds, corresponding to over $3\mbox{-}\sigma$ on either
side of the glitch epoch.
In the detection stage of the analysis, the prior on the effective amplitude is
chosen such that the probability density function is uniform across the
logarithm of the effective amplitude:
\begin{equation}\label{eq:amp prior}
p({\cal A}_{\rm eff}|M_+) = \frac{1}{\ln({\cal A}_{\rm eff}^{\rm upp}/{\cal A}_{\rm eff}^{\rm low}){\cal
A}_{\rm eff}}.
\end{equation}
This prior probability distribution is truncated at small (${\cal A}^{\rm
low}=10^{-22}$) and large (${\cal A}^{\rm upp}=10^{-19}$) values to ensure that
it is correctly normalized.  The lower truncation is chosen to be much smaller
than the effective amplitude produced by any detectable signal.  That is,
gravitational wave signals with effective amplitudes this small are
indistinguishable from detector noise and we do not benefit from extending this
lower limit.  Similarly, the upper truncation is chosen to be well above the
effective amplitude of easily detectable signals.  However, when we come to
form the posteriors on the amplitude and energy of gravitational waves we
instead adopt a uniform prior on the effective amplitude on the range
$[0,\infty)$, similar to the priors placed on frequency and decay time.

The reason for using these different priors in the different stages of the
analysis is that, in the first stage, we wish to weight lower amplitude signals
in keeping with astrophysical expectations and reduce the chance of falsely
identifying a loud instrumental transient as a \gw detection candidate.  By the
second stage, however, if we have already decided that there is no detection
candidate, we aim to set conservative upper limits on \gw amplitude and
energy without introducing any additional bias towards low amplitudes.  We find
that the logarithmically uniform amplitude prior lowers (strengthens) the
posterior amplitude upper limit from the uniform-amplitude case by
as much as $50\%$.  The linearly uniform amplitude prior is, therefore, more
appropriate for the construction of conservative upper limits.

%
The search described in this work uses data $D_1$ and $D_2$ from two detectors.
For the signal model, the data from each detector are combined by multiplying
the likelihood of $D_1$ with the likelihood of $D_2$ between the detectors:
\begin{eqnarray} 
p(D|\vec{\mu},M_+) & = &
p(D_1|\vec{\mu},M_+) p(D_2|\vec{\mu},M_+). \label{eq:joint detector
likelihood 1}
\end{eqnarray}
Notice that this expression assumes that the data streams are uncorrelated.  At
the frequencies of interest to this search (i.e. 1--3\,kHz), the dominant
source of noise is photon shot noise which is not be correlated between
detectors.  Studies in~\cite{S5HFburst} support this assumption.  In addition,
the frequentist interpretation of the odds ratios obtained from off-source
trials provides an additional level of robustness against common correlated
instrumental transient artifacts.

\subsection{Computing the evidence against gravitational wave detection}
We consider two possibilities which comprise the null-detection model: (i) 
Gaussian noise (model $N$) and (ii) an instrumental transient which is
uncorrelated between detectors (model $T$). For the noise model, there is no
contribution from any excess power due to \gws or instrumental transients.
The likelihood function for the full spectrogram is then given by the central
$\chi^2$ distribution with two degrees of freedom:
\begin{equation}
\label{eq:noise likelihood}
p(D|N) = \prod_{i=1}^{N_{\rm T}}\prod_{j=1}^{N_{\rm F}}
\frac{1}{2\sigma_j^2}e^{-\tilde{d}_{ij}/2\sigma_j^2}.
\end{equation}
%
A simple comparison between the signal model and $\chi^2$ distributed noise
is insufficient to discriminate real signals from instrumental
transients due simply to the fact that
\emph{any} excess power tends to resemble the signal model more closely than
the noise model.  Following~\citep{Veitch:2010}, we consider an alternative
scenario for null-detection in which there is a transient signal of environmental
or instrumental origin in the data.  This artifact can mimic the gravitational wave
ring-–down signal we expect from the pulsar glitch.
However, it may be present only in a single detector,
or there may be temporally coincident instrumental transients with
signal parameters inconsistent between detectors.  In this case, the data $D_1$
and $D_2$ are independent, so the evidence for model $T$ is simply the product
of the evidences in each data stream,
\begin{eqnarray}\label{eq:incoherent evidence}
\prob(D|T) = \prob (D_1|T) \prob (D_2|T). 
\end{eqnarray}
The individual evidences are computed according to
\begin{eqnarray}
\prob(D_i|T)  =  \int_{\vec{\mu}} p(\vec{\mu}|T) 
  p(D_i|\vec{\mu},T)~{\rm d}\vec{\mu}.
\end{eqnarray}
\subsection{Detection statistic \& upper limits}
The total evidence for the null-detection model is the sum of the evidence for
the instrumental transient model $T$ and the noise-only model $N$ and we are
left with the following expression for $\detodds$, our detection statistic:
\begin{equation}
\detodds = \frac{\prob(D|M_+)}{\prob(D|T) + \prob(D|N)}.
\end{equation}
Gravitational wave signals are correlated between detectors and, therefore,
lead to higher evidence for the detection model $M_+$ than the transient
model $T$.  The transient model is also penalized relative to the signal model
by virtue of the fact that the transient model has twice as many parameters over which
it is marginalized.  This yields a lower transient model evidence since it has been weighted
down by twice the number of prior probability distributions.  More importantly,
instrumental transients are generally uncorrelated between detectors.  If a
transient is only present in data stream $D_1$, for example, the likelihood from
data $D_2$ will be very small.  Multiplying these likelihoods inside the
evidence integral for the detection model leads to nearly zero overall evidence
for that model.  The transient model $T$, by contrast, does not suffer this
penalty so greatly since evidence may still be accumulated from other regions
of parameter space before the separate evidence integrals are multiplied.

In the absence of a detection candidate, we compute the marginal posterior
probability distribution on the effective amplitude ${\mathcal A}_{\rm eff}$,
directly from the data using the likelihood function in equation~\ref{eq:signal
likelihood} and the prior distributions discussed in the preceeding section.
This posterior is then transformed into three separate posteriors for each
value of $|m|$, according to Eq.~(\ref{eq:effective amplitude}).  These are
used to obtain Bayesian $90\%$ upper limits on the intrinsic strain amplitudes,
$h_{2m}$, by solving the following integral,
\begin{equation}\label{eq:amp UL}
0.9 = \int_0^{h_{2m}^{90\%}}p(h_{2m}|D,M_+)~{\rm d}h_{2m}.
\end{equation}
As described in Sec.~\ref{sec:theory}, we can use the expressions for the \gw
polarizations in Eqs.~(\ref{eq:hplus}) and~(\ref{eq:hcross}), to find the energy
emitted by \gws of different spherical harmonic modes by integrating the \gw
luminosity over solid angle and time.  The resulting expressions for the energy
from each harmonic all scale with the signal parameters
$\{h_{2m},~\nu_0,~\tau_0\}$  and distance to the source $\dist$ as,
\begin{equation}\label{eq:RD energy} 
E_{2m} \sim  \left(h_{2m}\nu_0\dist\right)^2\tau_0.
\end{equation}
The precise expression for each  harmonic includes a different numerical
factor, determined by integration of the ${\mathcal A}_{+,\times}^{2m}$ terms
over solid angle.  The relationships between the energies $E_{2m}$ and our
signal parameters allows us to form the marginal posterior probability density
for the energy from the $m^{\rm th}$ mode.  These energy posteriors can then be
used to find the energy upper limit by the same method as described above
for the \gw amplitude.

\section{Results}\label{sec:results}

\begin{figure}
\centering
\input{FAP_onsource.tex}
\includegraphics{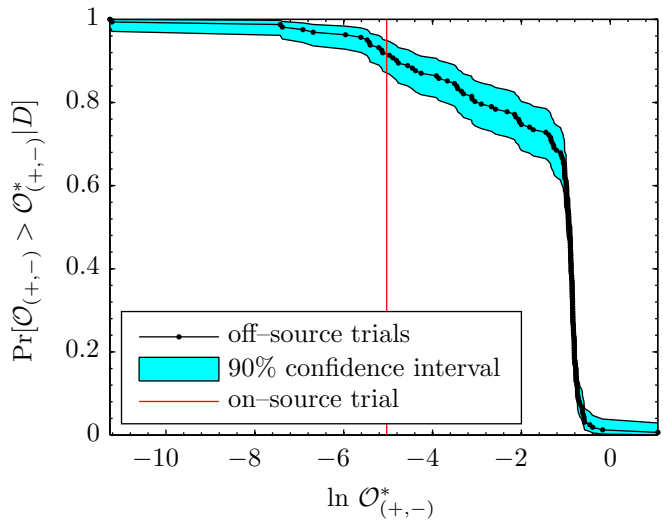}
\caption{The cumulative probability
distribution function (CDF) for the off-source value of $\ln \detodds$.  $\ln
\detodds^*$ indicates the observed value.  The shaded region shows the $90\%$
confidence interval on the estimate of the CDF.  The vertical line (red in the
on-line version)
indicates the value of $\ln \detodds = \onodds $ obtained from the
on-source data segment.  The probability of obtaining this
value or greater from background alone is $\FAP$, where the red line intersects
the black curve. The most significant off-source trial has $\ln \detodds =
\maxoffodds$ and the least significant has $\ln
\detodds=\minoffodds$.\label{fig:onsource FAPs}} \end{figure}

As stated in Sec.~\ref{sec:ligo data}, we have a total of 19586\,seconds of
completely contiguous H1 and H2 data for use in the analysis.  Our on-source
region is 120\,seconds centered on GPS time \glitchepochGPS.  This gives us
\preoffduration\,seconds of off-source data prior to  and
\postoffduration\,seconds of off-source data following the on-source region.
We assume that the noise characteristics of all of the off-source data remain
constant and are representative of the on-source.  We then split the
off-source data into segments of 120\,seconds to match the on-source region.
We obtain a maximum of 161 trials which can be used to estimate the
distribution of the odds ratio $\ln \detodds$ in the H1, H2 data.
Figure~\ref{fig:onsource FAPs} shows the cumulative distribution of $\ln
\detodds$ in the off-source data.  The largest value of the log odds found in
the 161 off-source trials is $\ln \detodds = \maxoffodds$.   The minimum value
is $\ln \detodds = \minoffodds$. Such a low value of the odds ratio indicates
that there is strong evidence in favour of the null-detection model and that
the data used for some this off-source trial contains one or more instrumental
transients inconsistent with \gw signals.  We set a threshold equal to the
loudest off-source value, above which we consider the on-source value to be
significant enough to merit further investigation.   The loudest off-source
value of $\ln \detodds = \maxoffodds$ corresponds to a false alarm probability
of 1/161.
%
The odds of the detection model versus the null-detection model in the
on-source data is $\ln \detodds = \onodds$, shown as the vertical line (red
in the on-line version)
in Fig.~\ref{fig:onsource FAPs}.  Using the results from the off-source
trials, we estimate that the probability of obtaining a value of $\detodds$
greater than the on-source value from background alone is $\FAP$.  We
therefore find no evidence in favour of \gw emission in the form of a
ring-down associated with this pulsar glitch.

\begin{figure} \input{energy_posterior.tex}
\includegraphics{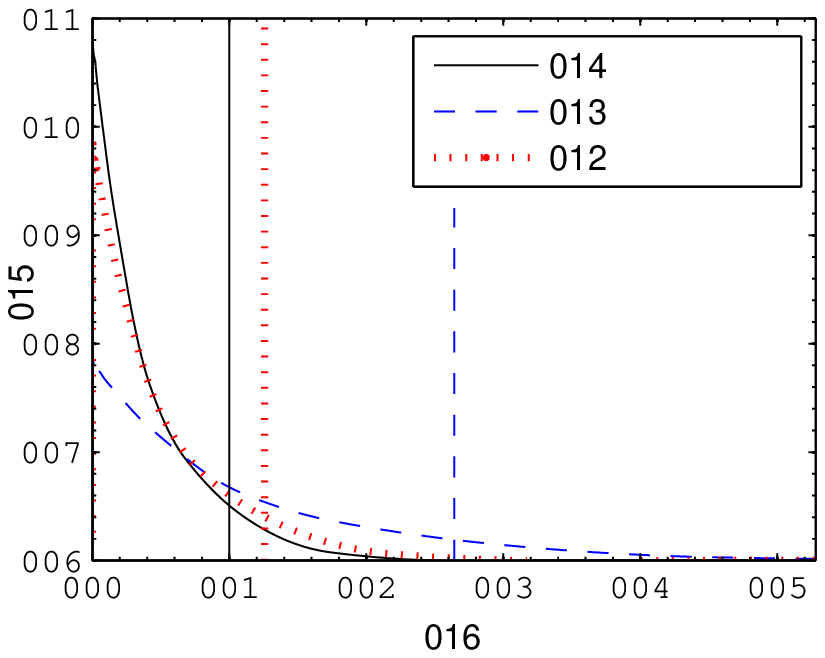} \caption{The posterior probability
density distributions and upper limits on the total \gw energy in the form of
ring-downs, assuming only a single harmonic (i.e. value of $|m|$) is excited.
The upper limits for each harmonic are shown as the vertical lines in the
figure.  The numerical values of the $90\%$ confidence upper limits can be
found in table~\ref{table:results}.  The $l=2,~|m|=0$ posterior is shown as the
solid black line, the dashed curve (blue in the on-line version) shows the $l=2,~|m|=1$ posterior and
the $l=2,~|m|=2$ posterior is shown as the dotted
curve (red in the on-line version).\label{fig:energy posterior}}\end{figure}

The marginal posterior probability distributions and $90\%$ confidence upper
limits on the peak intrinsic amplitude $h_{2m}$ and the total \gw energy
$E_{2m}$ for each value of $|m|$ are shown in Figs.~\ref{fig:amplitude
posterior} and~\ref{fig:energy posterior}.  The numerical values of the 
upper limits on amplitude and energy for different values of $|m|$ can be found
in table~\ref{table:results}.  We find that the different limits all lie within
a factor of $\sim 2$ of one another.  Note that these upper limits assume the
signal model $M_+$ is correct and, unlike the detection statistic $\detodds$,
do not directly account for instrumental
transients.  

During S5, the uncertainty in the magnitude of the detector response function
in the frequency band of interest was $\sim 15\%$ in H1 and $\sim 11\%$ in
H2~\cite{S5calibration} leading to uncertainties in the amplitude and energy
upper limits of $\sim 15\%$ and $\sim 30\%$, respectively.  Note that H1 is
the more sensitive detector and its calibration error dominates the analysis.


\begin{figure}
\centering
\input{amplitude_posterior.tex}
\includegraphics{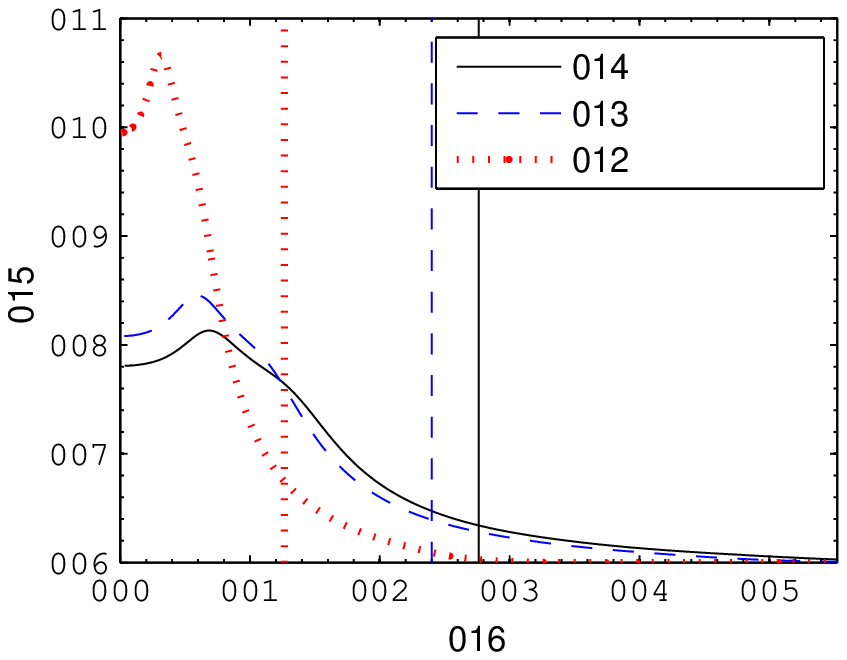}
\caption{The posterior probability density distributions and upper limits on
the intrinsic peak amplitude of ring-downs, assuming only a single harmonic
(i.e. value of $|m|$) is excited.  The upper limits for each harmonic are
shown as the vertical lines in the figure.  The numerical values of the $90\%$ confidence
upper limits can be found in table~\ref{table:results}.  The $l=2,~|m|=0$ posterior is shown as the solid
(black) line, the dashed curve (blue in the on-line version) shows the $l=2,~|m|=1$ posterior and the
$l=2,~|m|=2$ posterior is shown as the dotted curve (red in the on-line version).\label{fig:amplitude posterior}}\end{figure}

%

\begin{table}
\begin{center}
\begin{tabular}{l  c  c}
\hline Spherical Harmonic Indices  & $h_{2m}^{90\%} $ & $E_{2m}^{90\%}~\left({\rm erg}\right)$
\rule{0cm}{0.35cm}\\ \hline
$l=2,~m=0$ & $\htwozero$ & $\Etwozero$\rule{0cm}{0.35cm}\\
$l=2,~m=\pm1$ & $\htwoone$ & $\Etwoone$ \\
$l=2,~m=\pm2$ & $\htwotwo$ & $\Etwotwo$ \\
\end{tabular} \caption[]{The Bayesian $90\%$ confidence upper limits on the
intrinsic strain amplitude and energy associated with each spherical harmonic
mode of oscillation.\label{table:results}} 
\end{center} \end{table}

\section{Pipeline validation}\label{sec:validation}
The analysis pipeline is validated and its performance is characterized by
performing \emph{software injections} whereby a population of simulated signals with
parameters drawn from the prior distributions described in Sec.~\ref{sec:signal
model} are added to detector time-series data prior to
running the search algorithm.  We then count what fraction of the injection
population is recovered by the pipeline at increasing signal strengths.  This
fraction is the probability that a signal of a given strength will produce a
value of $\detodds$ larger than the largest off-source value, providing a
detection candidate.  

\begin{figure}\centering
\input{efficiency.tex} \includegraphics{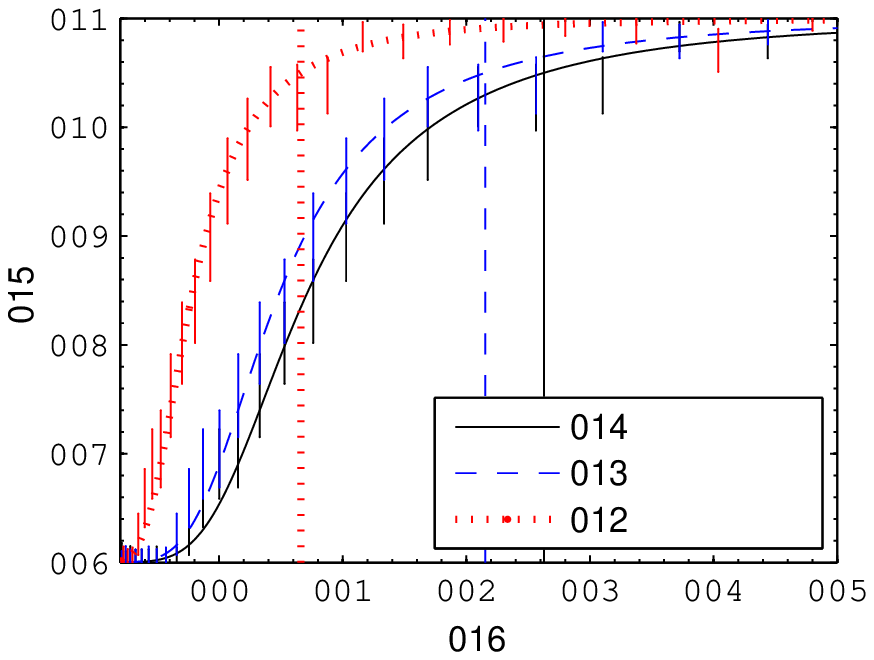} \caption{Detection
probabilities for software injections of ring-downs described by
Eq.~(\ref{eq:signal model}) and with parameters drawn from the prior
distributions used in the search and described in Sec.~\ref{sec:signal model}.
Efficiencies are computed by counting the number of detections in discrete bins
in amplitude.  The three different curves correspond to the different harmonics
(values of $|m|$).  The long vertical lines indicate the $90\%$ detection
efficiencies for the different harmonics, whose numerical values are given in
the text.  The shorter vertical bars indicate $90\%$ binomial confidence
intervals in the estimate of the detection probability in each amplitude
bin.  The color-coding and line-style convention is the same as that in
Figs.~\ref{fig:amplitude posterior} and~\ref{fig:energy posterior}: The
$l=2,~|m|=0$ efficiency curve is shown as the solid black line, the dashed curve
(blue in the on-line version)
shows the $l=2,~|m|=1$ efficiency curve and the $l=2,~|m|=2$ efficiency curve
is shown as the dotted curve (red in the on-line version).  \label{fig:efficiency}}
\end{figure}

Figure~\ref{fig:efficiency} shows the detection probabilities at increasing
values of initial intrinsic \gw amplitude for each harmonic mode.   A single
population of injections was generated with amplitudes drawn from the
logarithmically uniform prior on the effective amplitude ${\mathcal A}_{\rm
eff}$, given by Eq.~(\ref{eq:amp prior}).  Injection frequencies and damping
times are drawn from the uniform priors on those parameters.  Injection start
times, on the other hand, are drawn uniformly from both on- and off-source
times as a check to ensure that there is no bias in detection efficiency
between the on- and off-source data.  The three curves corresponding to the
harmonic modes $|m|=0,1,2$ are generated by scaling the effective amplitudes by
the detector antenna factors and appropriate inclination terms for each mode.
We characterize the sensitivity of the pipeline by the initial amplitude
required to reach $90\%$ detection probability.  These $90\%$ detection
efficiency amplitudes are marked in Fig.~\ref{fig:efficiency}.  For
$l=2,~|m|=0$, the $90\%$ efficiency amplitude is $h_{20}=1.8\times10^{-20}$;
$l=2,~|m|=1$ has $90\%$ detection efficiency at $h_{21}=1.6\times10^{-20}$ and
$l=2~|m|=2$ has $90\%$ detection efficiency at $8.3\times 10^{-21}$.  These are
$\sim 30\%$ larger than the corresponding Bayesian $90\%$ confidence upper
limits shown in Fig.~\ref{fig:amplitude posterior}.  This discrepancy is not
unexpected: the Bayesian amplitude posterior and the frequentist efficiency
curve ask entirely different questions of the data.  We therefore present the
efficiency curve purely as evidence that the analysis pipeline could have
detected a putative \gw signal, \emph{if there was one present}.  The Bayesian
upper limits, on the other hand, represent the strength of a \gw signal we
believe could have been present, \emph{given the on-source observations}.

\section{Discussion}\label{sec:discussion} 
We have performed a search for gravitational wave emission associated with a
timing glitch in PSR B0833$-$45, the Vela pulsar, during the fifth LIGO science
run. This search targeted ring-down signals in the frequency range
$[1,~3]$\,kHz, with damping times in the range $[50,~500]$\,ms.  No
gravitational wave detection candidate was found.  We place Bayesian $90\%$
confidence upper limits on the intrinsic peak \gw amplitude and total \gw
energy emitted by each quadrupolar spherical harmonic mode, assuming that only a
single mode dominates any neutron star oscillations associated with the glitch.
The amplitude and energy upper limits for the different modes are reported in
table~\ref{table:results}.  The upper limits for each value of $|m|$ agree with
one another to within a factor of $\sim 2$.  Investigations of the impact of
calibration uncertainties under a variety of scenarios suggest that the
uncertainties in these upper limits are no more than $\sim 15\%$ and $\sim
30\%$ in amplitude and energy, respectively.

Having presented upper limits on the gravitational wave emission for the
different possible values of the index $m$, we may ask what the
physical interest is in the different cases. In the absence of a
definitive model of the glitch mechanism, it is not possible to say in advance
which $m$ value is likely to be dominant.  However, the different
symmetries corresponding to the different $m$ values offers some insight into
the possible glitch mechanism.  The $|m|=0$ case corresponds to the excitation of
modes whose eigenfunctions are symmetric about the rotation axis.  In the
context of glitches, this would be rather natural, as glitches are thought to
be caused either by the build-up of a rotational lag (as in the superfluid
model) or by a build-up of elastic strain energy in response to  a decreasing
centrifugal force (the star-quake model); both of these are axisymmetric in
nature.  The $|m|=1$ case might correspond to a glitch that begins at one
point in the star before propagating outwards.  The $|m| = 2$ case might
correspond to a glitch that inherits the symmetry of the magnetic dipole field
that is believed to power the bulk of the star's spindown and radio pulsar
emission.    Clearly, a gravitational wave observation indicating which value
of $m$ (if any) of these is dominant will provide a unique insight into the
glitch mechanism.

It is natural to compare the upper limits presented here with other \gw
searches for $f$-mode ring-downs.  The only other search for a single
$f$-mode event~\cite{SGRsearch}
presents a best upper limit of $E^{90\%}_{\rm GW} = 2.4
\times10^{48}$\,erg on the energy emitted in \gws via $f$-mode induced
ring-down signals associated with a flare from SGR~1806$-$20 on UTC 2006--08--24
14:55:26.  In that analysis, however, the upper limits  assume isotropic
\gw emission.  In addition, the nominal distance of SGR~1806$-$20 is 10\,kpc.
To compare our results with those in~\cite{SGRsearch}, we must rescale our
upper limit on the effective amplitude ${\mathcal A}_{\rm eff}$ to a source distance of
10\,kpc,  assume isotropic \gw emission and use the average antenna factor of
$(F_+^2 + F_{\times}^2)^{1/2}=0.3$.  We then find our equivalent, isotropic
energy upper limit to be $1.3\times 10^{48}$\,erg, a factor of $\sim
2$ lower than that in~\cite{SGRsearch}.  This improvement is to be expected
since the analysis presented here assumes that the signal
waveform is a decaying sinusoid.  The analysis in~\cite{SGRsearch}, by contrast, 
does not rely on a particular waveform and is designed to search for bursts of excess power with
durations and frequencies compatible with $f$-mode ring-down signals.

Following the arguments laid out in Sec.~\ref{sec:theory}, the characteristic
energy of a pulsar glitch is believed to be of order $10^{38}$ or
$10^{42}$\,erg, depending on the mechanism.  Our current energy upper limits
are 2--3 orders of magnitude above (weaker than) the more optimistic theoretical limit.
The next generation of \gw observatories currently under construction, such as advanced LIGO~\citep{advLIGO}
and advanced Virgo~\cite{aVirgo}, is expected to have noise amplitude more than an order of
magnitude lower than in the current LIGO detectors at $f$-mode frequencies.
This corresponds to probing energies more than two orders of magnitude lower
than is currently possible, comparable to the order $10^{42}$\,erg of the most
optimistic theoretical predictions.
The detection of \gws associated with a Vela
glitch in the advanced interferometer era is therefore possible and would
provide compelling observational evidence for the star-quake theory of pulsar
glitches.
According to current conceptual design~\cite{ET}, the planned Einstein
Telescope would improve noise amplitude at $f$-mode frequencies another order of magnitude beyond
advanced LIGO, thereby improving the Vela glitch energy sensitivity two orders
of magnitude to of order $10^{40}$\,erg.

\begin{acknowledgments} The authors gratefully acknowledge the support of the
United States National Science Foundation for the construction and operation of
the LIGO Laboratory and the Science and Technology Facilities Council of the
United Kingdom, the Max-Planck-Society, and the State of Niedersachsen/Germany
for support of the construction and operation of the GEO\,600 detector.  The
authors also gratefully acknowledge the support of the research by these
agencies and by the Australian Research Council, the Council of Scientific and
Industrial Research of India, the Istituto Nazionale di Fisica Nucleare of
Italy, the Spanish Ministerio de Educaci\'on y Ciencia, the Conselleria
d'Economia Hisenda i Innovaci\'o of the Govern de les Illes Balears, the Royal
Society, the Scottish Funding Council,  the Scottish Universities Physics
Alliance, The National Aeronautics and Space Administration, the Carnegie
Trust, the Leverhulme Trust, the David and Lucile Packard Foundation, the
Research Corporation, and the Alfred P. Sloan Foundation.
This paper has been assigned LIGO Document No.\ P1000030-v11.
\end{acknowledgments}

\bibliography{Vela_Glitch}

\end{document}